\documentclass[reprint,showpacs,preprintnumbers,amssymb,superscriptaddress,aps,prd,nofootinbib]{revtex4-1}

\usepackage{graphicx, epsfig, amssymb} 
\usepackage{amsmath, amsfonts}
\usepackage{bm}

\usepackage[linktocpage]{hyperref}
\usepackage[usenames]{color}
\usepackage{subfigure}
\usepackage[utf8]{inputenc}
\usepackage{tensor}
\usepackage{soul}

\def\b{\begin{equation}}
\def\e{\end{equation}}

\def\be{\begin{eqnarray}}
\def\ee{\end{eqnarray}}

\begin{document}\title{\large Addendum to ``Absorption of a massive scalar field by a charged black hole''}

\author{Carolina L. Benone}\email{lben.carol@gmail.com}
\affiliation{Faculdade de F\'{\i}sica, Universidade Federal do Par\'a, 66075-110, Bel\'em, Par\'a, Brazil}

\author{Ednilton S. de Oliveira}\email{esdeoliveira@gmail.com}
\affiliation{Faculdade de F\'{\i}sica, Universidade Federal do Par\'a, 66075-110, Bel\'em, Par\'a, Brazil}

\author{Sam R. Dolan}\email{s.dolan@sheffield.ac.uk}
\affiliation{Consortium for Fundamental Physics,
School of Mathematics and Statistics,
University of Sheffield, Hicks Building, Hounsfield Road, Sheffield S3 7RH, United Kingdom}

\author{Lu\'{\i}s C. B. Crispino}\email{crispino@ufpa.br}
\affiliation{Faculdade de F\'{\i}sica, Universidade Federal do Par\'a, 66075-110, Bel\'em, Par\'a, Brazil}

\begin{abstract}
In Phys.~Rev.~D{\bf 89}, 104053 (2014) we studied the absorption cross section 
of a scalar field of mass $m$ impinging on a static black hole of mass $M$ and charge $Q$. We presented
numerical results using the partial-wave method, and analytical results in the high- and low-frequency limit. 
 Our low-frequency approximation was only valid if the (dimensionless) field velocity
$v$ exceeds $v_c = 2 \pi M m$. In this Addendum we give the complementary result for $v \lesssim v_c$, and we consider the possible physical relevance of this regime.
\end{abstract}

\date{\today}

\maketitle

In Ref.~\cite{bodc} we analyzed the scenario of a neutral scalar field of mass $m$ and frequency $\omega$ 
absorbed by a Reissner-Nordstr\"om black hole of mass $M$ and charge $Q$. Our stated aim was to provide a quantitative full-spectrum description of absorption, by bringing together numerical methods and analytical approximations. However, due to a tacit assumption, we gave an incomplete description of the low-frequency regime; with this addendum we now fulfill our original objective.

The total absorption cross section $\sigma$ may be written as a sum of partial absorption cross sections $\sigma_l$. In the low-frequency limit $M \omega \ll 1$, the dominant contribution arises from the monopole sector, $\sigma_{lf} \approx \sigma_{l=0}$. In Ref.~\cite{bodc} we obtained $\sigma_{lf}=\mathcal{A} / v$ [see Eq.~(63)], where $v=\sqrt{1-m^2/\omega^2}$ is the (dimensionless) velocity of the field, $\mathcal{A}=4\pi r_+^2$ is the area of the black hole, and $r_+$ is the areal coordinate of the event horizon location. 

However, our original result did not encompass the limit of small velocities. After taking this into account, the completed low-frequency approximation is
\be
\sigma_{lf} = \begin{cases} 
\sigma_{lf}^{(1)} = \mathcal{A} / v , & v \gtrsim v_c , \\
\sigma_{lf}^{(2)} = \frac{4(\pi r_+)^2(2 M m)}{v^2} , & v \lesssim v_c ,
\end{cases}
\label{lf1}
\ee
where $v_c = 2\pi Mm$ is the velocity of the transition.

One may obtain the approximation valid 
for $v \lesssim v_c$ by
starting from Eq.~(62) of Ref.~\cite{bodc}, namely
\b
\sigma = \frac{ 4\pi r_+^2 \rho^2 }{v},
\label{sigmarhov}
\e
where we are considering only the first term in the denominator 
of Eq. (62). 
For this purpose, we write~\footnote{There are $2\pi$ factors (in front of $\eta$) 
missing in Eq. (57) of Ref.~\cite{bodc}. 
}
\b
\rho^2
=\frac{2\pi\eta}{e^{2\pi\eta}-1}\\
\label{rho}
\e
as
\b
\rho^2
=
-\frac{2\pi Mm(1+v^2)}{v \sqrt{1-v^2}}
\frac{1}{\exp\left(-\frac{2\pi Mm(1+v^2)}{v \sqrt{1-v^2}}\right)-1}.
\label{rhov}
\e
%
%
Substituting Eq.~(\ref{rhov}) in 
Eq.~(\ref{sigmarhov}) 
and considering the limit for $v\rightarrow 0$ we obtain
\b
\sigma_{lf}^{(2)} = \frac{4(\pi r_+)^2(2 M m)}{v^2}.
\label{lf2}
\e
In the uncharged case $Q=0$ we recover Unruh's 
result for the case of a Schwarzschild black hole 
(cf. Eq. (97) of Ref.~\cite{Unruh_1976}). 

\begin{figure}[h!]
\includegraphics[width=\columnwidth]{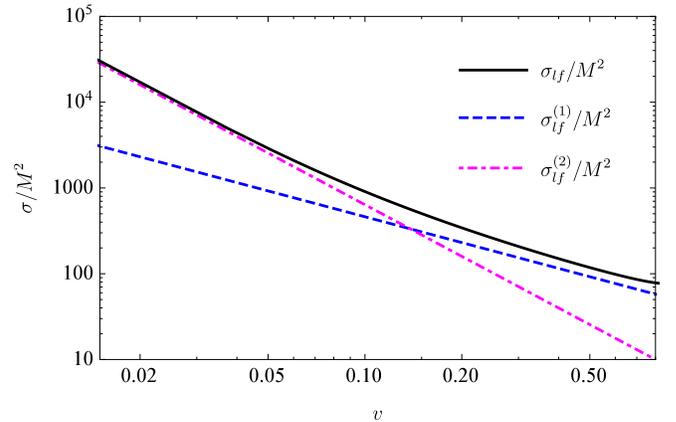}
\caption{
Comparison between the partial absorption cross section $\sigma_{l=0}$ [solid line], 
and the approximate analytical results of Eq.~(\ref{lf1}) [broken lines], for the case 
$Q/M=0.4$ and $Mm=0.022$. A transition in behaviour is visible near $v_c \approx 0.138$.}
\label{fig1}
\end{figure}

In Fig.~\ref{fig1} we compare Eq.~(\ref{lf1}) with our numerical results.   
We can see that the numerical results present a transition between $\sigma_{lf}^{(1)}$ and $\sigma_{lf}^{(2)}$, 
which happens near $v=v_c$, with $v_c \approx 0.138$ in this case.

Recently a new dark matter candidate has been proposed by Hui \emph{et al.}~\cite{Hui:2016ltb}, in the form of a scalar field with mass $m \approx 10^{-22} \, \text{eV} / c^2$ and de Broglie wavelength $\lambda_B \approx 1 \, \text{kpc}$. Its corresponding velocity is $v \approx 4 \times 10^{-4}$, found from $v = \left(1 + \lambda_B^2 / \lambda_{C}^2 \right)^{-1/2}$, where $\lambda_C = h / m c \approx 0.4 \, \text{pc}$ is the Compton wavelength. For a black hole mass $M_1 = 3.6 \times 10^6 M_\odot$ (e.g.~Sgr.~A* \cite{Schoedel:2009mv}), we find $v_c \approx 1.7 \times 10^{-5}$ and thus $v > v_c$; whereas for a supermassive black hole of mass $M_2 = 2 \times 10^8 M_\odot$ (e.g.~Andromeda's supermassive BH has mass $(1.1-2.3) \times 10^8 M_\odot$ \cite{Bender:2005rq}) we find $v_c = 9.4 \times 10^{-4}$ and thus $v < v_c$. This suggests that both regimes of Eq.~(\ref{lf1}) are potentially relevant in the scenario of Hui \emph{et al}., and that disparate cross sections are possible. For example, $\sigma_{lf}^{(1)} \approx 3.6 \times 10^{24} \text{m}^2$ for $M_1$ (Sgr.~A*) and $\sigma_{lf}^{(2)} \approx 2.6 \times 10^{28} \text{m}^2$ for $M_2$ (Andromeda). 

\acknowledgments
We thank E.~Witten for raising a question by email that led to this Addendum. 
We acknowledge Conselho Nacional de Desenvolvimento
Cient\'ifico e Tecnol\'ogico (CNPq), Coordena\c{c}\~ao de 
Aperfei\c{c}oamento de Pessoal de N\'ivel Superior (CAPES),  
Marie Curie action NRHEP-295189- FP7-PEOPLE-2011-IRSES, 
Engineering and Physical Sciences Research Council (EPSRC) Grant No.~EP/M025802/1 
and Science and Technology Facilities Council (STFC) Grant No.~ST/L000520/1 for partial financial support.

\end{document}